\documentclass[conference, 10pt]{IEEEtran}
\IEEEoverridecommandlockouts
\usepackage{cite}
\usepackage{amsmath,amssymb,amsfonts}
\usepackage{algorithmic}
\usepackage{graphicx}
\usepackage{textcomp}
\usepackage{caption}
\usepackage{subcaption}
\usepackage{multirow}
\usepackage{graphicx}
\usepackage[table,xcdraw]{xcolor}
\usepackage{xurl}
\usepackage[colorlinks=false]{hyperref}
\usepackage{comment}
\usepackage{xspace}

\def\BibTeX{{\rm B\kern-.05em{\sc i\kern-.025em b}\kern-.08em
    T\kern-.1667em\lower.7ex\hbox{E}\kern-.125emX}}

\newcommand{\FC}{\textsc{fc}\xspace}
\newcommand{\CONV}{\textsc{conv}\xspace}
\begin{document}


\title{Improving inference time in multi-TPU systems with profiled model segmentation}

\author{\IEEEauthorblockN{Jorge Villarrubia, Luis Costero, Francisco D. Igual, Katzalin Olcoz}
\IEEEauthorblockA{\textit{Dept. Arquitectura de Computadores y Autom\'atica} \\
\textit{Universidad Complutense de Madrid}\\
Madrid, Spain\\
\{jorvil01,lcostero,figual,katzalin\}@ucm.es}
\vspace*{-25pt} 
}

\maketitle

\begin{abstract}
In this paper, we systematically evaluate the inference
performance of the Edge TPU by Google for neural networks
with different characteristics. Specifically, we determine
that, given the limited amount of on-chip memory on the
Edge TPU, accesses to external (host) memory rapidly become
an important performance bottleneck. We demonstrate how
multiple devices can be jointly used to
alleviate the bottleneck introduced by accessing the host
memory. We propose a solution combining model segmentation
and pipelining on up to four TPUs, with remarkable performance
improvements that range from
$6\times$ for neural networks with convolutional layers to $46\times$
for fully connected layers, compared with single-TPU setups.
\end{abstract}

\begin{IEEEkeywords}
Domain-specific architectures, Edge TPU, deep learning, model segmentation.
\end{IEEEkeywords}

\section{Introduction}\label{sec:intro}

The increasing computational demands of the Internet of Things (IoT) applications, specifically in the
field of Artificial Intelligence (AI), and their sensitivity to latency and response time, have inspired the convergence of Edge
Computing and AI, moving the computation of some AI tasks closer to the sensors. Edge-AI~\cite{Li2019,Ren2022}
aims at executing AI algorithms
in power-constrained low-performance edge devices, 
in order to reduce latencies, increase security or alleviate the load of datacenters.
Edge-AI tasks typically include the execution of neural networks for a plethora of applications, including
object detection for smart cameras~\cite{James2019}, 
smart city applications~\cite{Thalluri2021}, healthcare~\cite{Alshehri2021ACS} or autonomous driving~\cite{McEnroe2022}, 
among others.

However, devices that operate on the edge usually suffer from a lack of performance that limit their exploitation
for compute-intensive tasks, together with severe energy restrictions. 
Combined, their use for application acceleration and the selection of the most appropriate
architecture for a specific problem or scenario is still an open challenge.
The use of multi-core CPUs, or general-purpose accelerators such as GPUs can alleviate the performance
problem, but they still exhibit non-negligible power consumption that, in
many situations, does not make up for the performance benefits.
The emergence of domain-specific architectures (DSAs) in the form of ASICs (Application Specific Integrated Circuits),
aims at alleviating this problem. Modern DSAs for Machine Learning (ML), such as
the Intel NCS~\cite{movidius_tech}, or the Google Edge TPU~\cite{coral_tech}, have been recently
introduced as an attractive trade-off between performance, energy efficiency and
flexibility for Edge-AI. Their generality and applicability to accelerate any
Deep Learning (DL) model remains as an open question.

In this work, we focus on one of those ASICs: the Edge TPU in
its PCIe version, and we propose a detailed performance study
hosting DL tasks. Specifically, we perform
a parametric evaluation based on the workload (size) of
a number of synthetic DL models, identifying the main
bottlenecks of using a single TPU for inference, and
alleviating them by introducing model segmentation techniques
that distribute the model across multiple TPUs.
Our contributions can be summarized as:

\begin{itemize}
\item We systematically evaluate the inference performance of single-TPU systems by evaluating synthetic models with both fully-connected (\FC) and convolutional (\CONV) layers with increasing computatitonal requirements.
\item We provide evidences and gain insights into the performance bottleneck introduced by the use of host memory for weight storage, specifically for large models that do not fit on the internal device memory.
\item We propose profiling-based model segmentation techniques to overcome the memory limitations of single-TPU systems, mapping the model to multi-TPU architectures in a pipelined fashion, with remarkable performance benefits and speedups on up to 4 TPUs ranging from 
$6\times$ for \CONV models
to
$46\times$ for \FC models, taking single-TPU executions as the baseline.
\end{itemize}

The rest of the paper is structured as follows.
Section~\ref{sec:background} provides a description of the Edge TPU architecture, and an overview of the state of the art in performance evaluation of TPUs.
Section~\ref{sec:macops} provides a parametric performance evaluation on a single TPU, identifying the main bottlenecks of the architecture.
Section~\ref{sec:memory} delves into these limitations in terms of memory occupation derived by the limited amount of on-chip memory, and its implications in terms of performance.
Section~\ref{sec:multitpu} proposes and evaluates a mechanism to overcome the previous limitations by distributing the inference to multiple TPUs.
Section~\ref{sec:conclusions} closes the paper with some final remarks.
\section{Background}\label{sec:background}

\subsection{The Edge TPU architecture}\label{sec:background:architecture}

\begin{figure}[ht!]
    \centering
    \centering
    \includegraphics[width=\columnwidth]{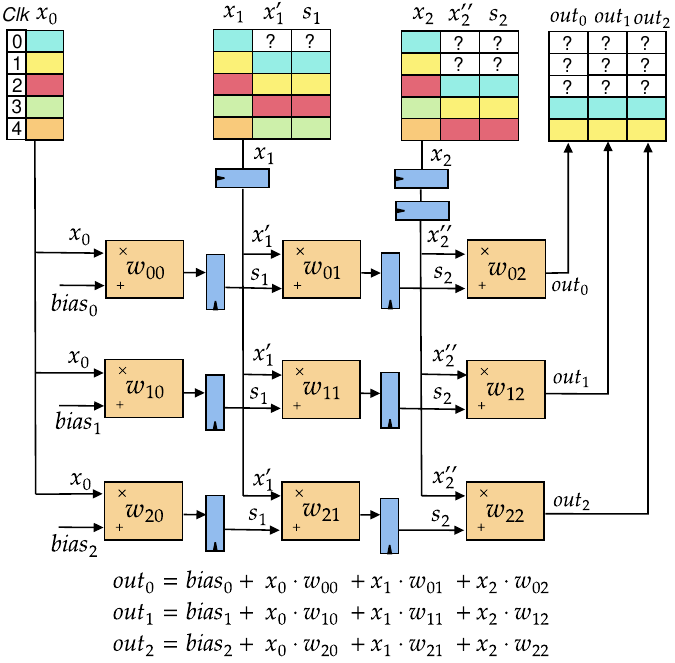}
    \caption{Example of a 3x3 systolic matrix and the cycle-by-cycle data flow through the chains.}
    \label{fig:systolic_matrix_example}
\end{figure}

The main inference operation in a neural network is the scalar product between input vectors and weight vectors (other operations such as evaluating on the activation function are much less expensive). To speed them up, TPUs include chains of multiply-sum cells~\cite{coral_tech}. In each cell, the product of a weight is calculated by its corresponding component of an input vector, the result is added to the cumulative product of the previous components (received from the previous cell) and propagated forward. These chains are segmented by registers so that the products of different input vectors can be run in parallel (several inferences can be made with the same neural network at the same time). When the size of the vectors exceeds the size of the chain, they are divided into fragments whose scalar products can also be calculated in parallel within a chain (in this case, the partial accumulations of each fragment are reduced to a single scalar at the output of the chain). In addition, the same inputs can be multiplied simultaneously by other weight vectors in other chains. For example, each chain can perform the product of the inputs of a layer by the weights of a neuron and several chains can simultaneously compute several neurons. This multi-chain structure constitutes a matrix of cells known as ``systolic matrix'' due to the thrust of the data with the clock pulse (similar to the systolic thrust of the blood in the heart). Figure~\ref{fig:systolic_matrix_example} shows an example of a $3\times3$ systolic matrix that calculates the scalar products of several inputs $(x_0, x_1, x_2)$ (each identified with a colour) by the weights $(w_{i0}, w_{i1}, w_{i2})$ of 3 neurons $n_i$ ($i \in \{0,1,2\}$). The propagation of each input cycle by cycle through the chains is indicated by the colours.

Although Google has not disclosed some of the Edge TPU's specifications, there are well-founded estimates that it incorporates a $64\times64$ systolic matrix running at a maximum frequency of 480 MHz. This is consistent with the 4 TOPS peak performance specified in its datasheet ($64 \cdot 64 \text{ cells/cycle } \cdot 2 \text{ ops/cell } \cdot 480 \cdot 10^6 \text{ cycles/s } \simeq 4 \cdot 10^{12} \text{ ops/s}$). In addition, as it is aimed at low power consumption environments, the device uses only 2W of power at maximum performance, offering an energy efficiency of up to 2 TOPS/W. This efficiency is achieved because all calculations in the systolic array are performed with integer arithmetic and multiplications are done with reduced 8-bit precision. This has many benefits in terms of performance, power consumption and hardware cost, but means that the chip can only be used for inference (higher precision is often required for training). In addition, a quantization process is needed before inference to transform the weights of the model trained with \emph{float32} to \emph{int8}.

On the other hand, the Edge TPU also includes an internal memory of 8 MiB where it stores the instructions (CISC repertoire of very high abstraction level), the inputs/activations and the weights of the model. In our setup, the chip is embedded in an M.2 module that connects via PCIe to a host system that invokes it for inference by providing the instructions and data. The host uses the \emph{edgetpu compiler} to generate the code and adapt the model operators to those implemented in the TPU; the models must depart from a TFLite model, and be previously quantized to 8-bit integers.


%
%

\subsection{Related work}\label{sec:background:relatedwork}

Since its introduction in 2019, both the Cloud and Edge versions of the TPU have received a significant attention in the literature, mainly assessing their performance and energy efficiency for model training and inference, respectively.

Several papers have analyzed the cloud version of the TPU. Specifically, \cite{Nikolic2022, Raj2020} provide an in-depth survey of the differences and particularities of CPUs, GPUs and (Cloud) TPUs for different tasks related with deep learning. The designers of the ASIC presented in \cite{Jouppi2017} a detailed description of the architecture, evaluating its performance and energy efficiency in comparison with other state-of-the-art architectures.

The Edge version of the TPU has received less attention in the literature~\cite{Cass2019}. \cite{Murshed2022} considers the Edge TPU as one of the testsbeds to evaluate ASICs for image processing tasks. \cite{Kang2022} evaluates the Edge TPU for object detection activities at the edge. \cite{Antonini2019} provides a comparative study of different edge accelerators, including the Edge TPU, for personal sensing applications. \cite{Seshadri2021,Sun2021} generalize the study by covering a number of DL models for the performance evaluation of the Edge TPU. The previous works, however, provide insights for specific (not synthetic) deep neural networks and/or applications, without further insights for generalizing the conclusions to other (existing or non-existing) models; none of them covers the use of multiple TPUs in the evaluation.

The Edge TPU has also been included as a member of the family of devices
capable to accelerate Edge-AI tasks by means of benchmarks. \cite{Varghese2021} includes the device as one of the target architectures appealing for state-of-the-art on edge performance benchmarks. MLPerf is an initiative to design portable benchmarks and to evaluate them on different architectures. MLPerf defines a set of models that are evaluated for training (mainly using high performance architectures, e.g. GPUs) and also for inferencing. In the latter case, {\em MLPerf Inference} includes different variants for datacenter, mobile, tiny and edge devices. Although the Edge TPU has not been included in any results for the edge MLPerf inference benchmarks, \cite{Libutti2020} performed a detailed study of its performance for a subset of the models used in the benchmark for the USB version of the device. 
\section{Inference performance on a single TPU}\label{sec:macops}

As previously stated, the inference in a neural network is based on dot products, carried out in the cells of the systolic matrix via multiply-accumulate (MAC) operations. Hence, it seems natural that the number of MAC operations is intimately related with the inference time. Our first round of experiments evaluates the behavior of the TPU when executing models with an increasing number of MAC operations. For that, we systematically generate and  evaluate \FC and \CONV models.  

\subsection{Synthetic model generation}

The \FC models are generated with $L_{FC}$ layers, varying the number of nodes $n$ for each layer in the range $[N_{min}, N_{max}]$ with step $S_{N}$. For the \CONV models, we deploy $L_{CONV}$ convolutional layers varying the number of filters $f$ for each one in the range $[F_{min}, F_{max}]$ with step $S_{F}$.

In the \FC layers, each weight is multiplied and accumulated exactly once, and hence the number of MAC operations matches the amount of weights of the network\footnote{The {\em bias} operations are ignored as they are not based on MAC; anyway, they grow linearly with $n$, and hence their impact is asymptotically negligible compared with the $n^2$ MAC operations per \FC layer.}. For the \CONV layers, the convolution filters are applied in a stride 1 fashion. %
Hence, each weight is multiplied and accumulated once per input image element; the processing of the input layer requires $C \cdot W \cdot H \cdot f \cdot F_w \cdot F_h$ MAC operations (where $C$ is the number of input channels, $W \times H$ the dimensions of each channel and $F_w \times F_h$ the dimensions of each filter). For the remaining layers, the number of input channels matches the number of filters $f$ of the previous layer, yielding $f \cdot W \cdot H \cdot f \cdot F_w \cdot F_h$ MAC operations. Summarizing, the number of MAC operations for a given layer with $f$ filters results $\#\text{MACs}\,(f) =  W \cdot H \cdot f \cdot F_w \cdot F_h \cdot \left(C + f \cdot (L_{\text{CONV}}-1)\right)$, that grows linearly with $f$ if $L_{\text{CONV}}= 1$ or quadratically if $L_{\text{CONV}} > 1$.

\subsection{Performance evaluation}

Armed with the previous data, the number of MAC operations for different models generated varying $n$ or $f$ can be easily calculated. Let us start by evaluating the performance of a single Edge TPU for \FC and \CONV models increasing the number of MAC operations.

\begin{figure*}[t]
  \begin{subfigure}[b]{0.32\textwidth}
    \centering
        \includegraphics[width=\textwidth]{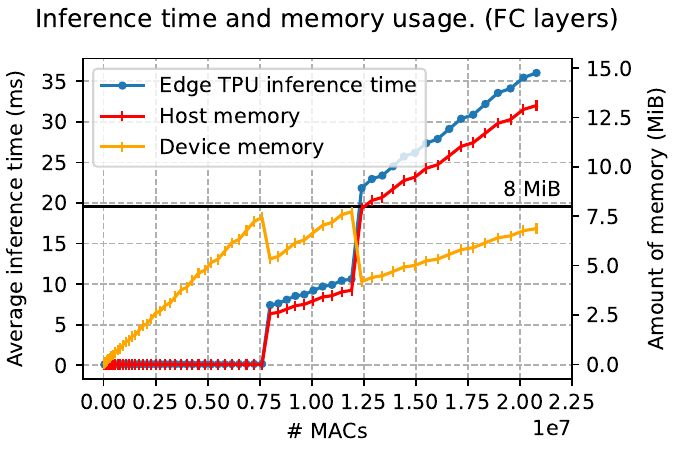}
        \includegraphics[width=\textwidth]{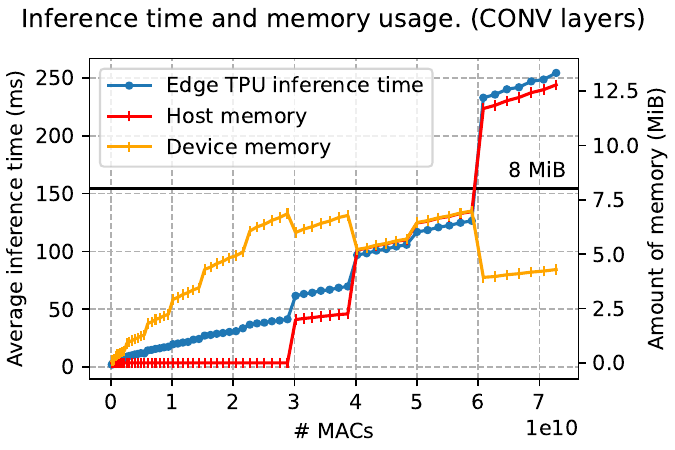}
    \caption{Inference time and memory usage for \FC (top) and \CONV (bottom) models.}
    \label{fig:FC_CONV_time_memory_MAC}
  \end{subfigure}
  \hfill
  \begin{subfigure}[b]{0.32\textwidth}
    \centering
        \includegraphics[width=\textwidth]{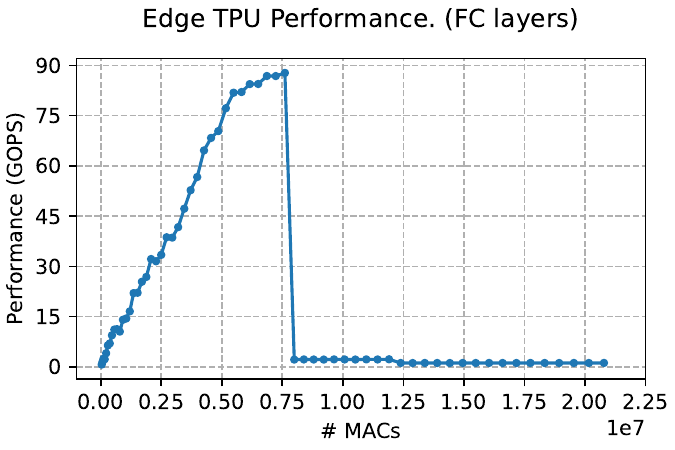}
        \includegraphics[width=\textwidth]{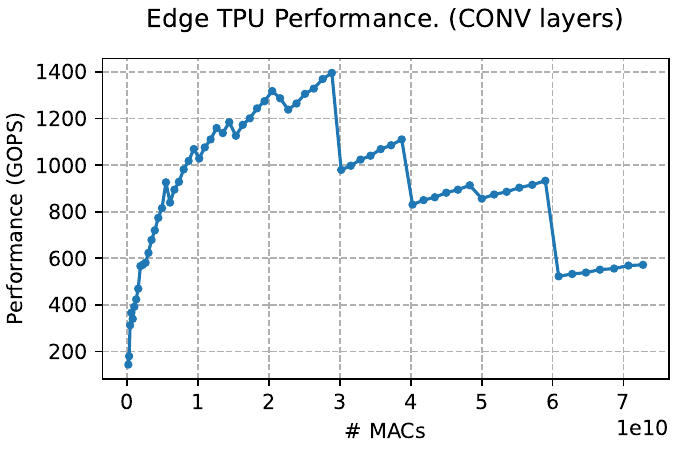}
    \caption{Performance for \FC (top) and \CONV (bottom) models.}
    \label{fig:FC_CONV_performance_MAC}
  \end{subfigure}
  \hfill
  \begin{subfigure}[b]{0.32\textwidth}
    \centering
        \includegraphics[width=\textwidth]{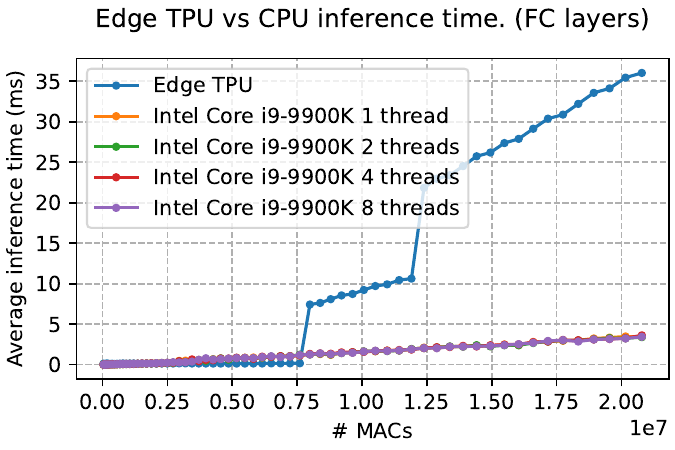}
        \includegraphics[width=\textwidth]{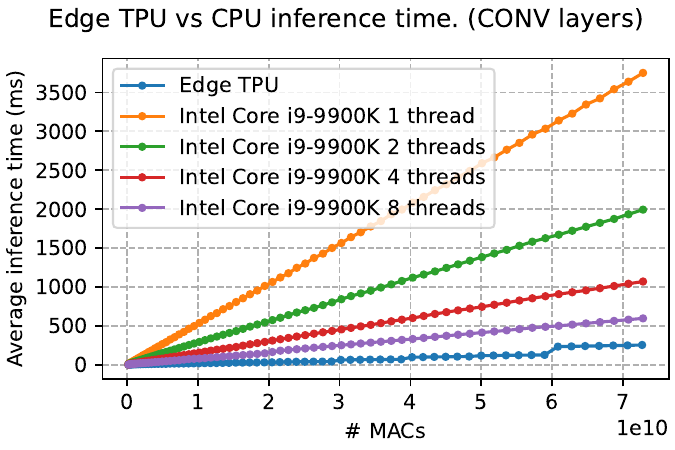}
    \caption{Inference time comparison between CPU and Edge TPU.}
    \label{fig:EdgeTPUvsCPU_time}
  \end{subfigure}
  \caption{Inference time analysis for a single Edge TPU.}
\end{figure*}

The blue lines in Figure~\ref{fig:FC_CONV_time_memory_MAC} report the average inference time for each model type; Figure~\ref{fig:FC_CONV_performance_MAC} reports the corresponding performance in terms of GOPS (billions of MACs per second). For the \FC layers, the models were generated varying the number of nodes per layer ($n$) with the following layer configuration: $L_{\text{FC}} = 5$, $I= 64$, $O = 10$, $N_{min} = 100$, $N_{max} = 2640$ and $S_N = 40$. For the \CONV layers, we varied the number of filters per layer $f$ with the following layer configuration: $L_{\text{CONV}} = 5$, $C = 3$, $W \times H = 64 \times 64$, $F_h\times F_w = 3 \times 3$, $F_{min} = 32$, $F_{max} = 702$ and $S_F = 10$.

A number of insights can be extracted from the previous results, that
motivate the experiments in the next sections:

\begin{itemize}

\item For both types of layers there is an evident stepped behaviorthat reveals dramatic increases in inference time for models with similar number of MAC operations. These leaps in inference time (that will be characterized and reduced throughout the following sections) arise between zones in which the inference time evolves at a much lower pace with respect to the number of operations; actually, mild performance increases can be observed in these areas as the number of MAC operations increase. 

\item The attained performance is dramatically lower than the theoretical peak of the architecture (4 TOPS), and differs across layer types. The executions are actually limited by memory accesses (they are {\em memory bound} for the range of MAC operations tested), and the lower arithmetic intensity of the \FC versus the \CONV layers explains the relative difference in performance between both types of layers.

\item The performance obtained by \CONV layers is much higher than that of the \FC layers; as an example, the peak performance in terms of GOPS for \CONV models is $17\times$ than that of a \FC model. As previously stated, the weights in \FC layers are used solely in a MAC operation; \CONV weights are reused in a number of operations (as they move through the input matrices), and hence memory movements can be amortized with computations.

\end{itemize}
\section{Analysis of host and device memory usage}\label{sec:memory}

Motivated by the previous observations, we proceed by analyzing the executions in terms of memory accesses. As of today, there are no profiling tools to obtain this type of metrics; the model compiler, however, generates a compilation report that includes the amount of host and device memory used by the Edge TPU to store the weights of the model. This information is a good indicator of the cost associated to memory operations, as the read operations associated to weights are the dominant operations in the inference process in terms of execution time (compared with the read or write of inputs and outputs, respectively, as the associated tensors in those operations are smaller). Additionally, the communications between host and TPU (in our case, via the PCIe bus) are a non-negligible bottleneck that should be reduced or avoided. In situations in which the model weights do not fit in device memory, the compiler will necessarily divide the storage of weights between device and host memory. The reduced size of the device memory (8 MiB) makes it common to find models that, even after quantization, cannot be completely stored in the device memory throughout the computation.

To support our discussion, Figure~\ref{fig:FC_CONV_time_memory_MAC} adds the amount of device and host memory used by the models deployed in the previous section. The memory usage perfectly explains the stepped behavior of the inference time. At each step, the device memory usage increases progressively until it reaches the available memory (8 MiB). Then, it drops sharply coinciding with an increase of host memory usage. This happens when part of the model cannot be stored on the device and starts being stored on the host. The overhead of loading these weights from the host causes the increase in inference time.

The host memory usage occurs in abrupt steps because the neural layer is the minimum storage unit: the Edge TPU compiler stores all the weights of a layer in the same memory. Theoretically, the tensors could be divided to store only the strictly necessary part of the model on the host; the compiler, however, proceeds by storing complete tensors, presumably for a simpler weight management. Their solution should perform the same number of memory copies than a storage scheme with a finer granularity, but with more data in each one. 

Table~\ref{tab:FC_mem_steps} reveals that the host memory usage becomes approximately half of device memory usage for \FC layers after the first step. The reason is that one of the three large layers of the model starts to be saved on the host and there are only two large layers left on the device (there are three hidden layers with $n^2$ weights that make the $64n$ weights of the first layer and the $10n$ weights of the output layer insignificant). Similarly, a second large layer is stored on the host after the second step leaving only one big layer on the device (thus host memory usage is about twice that of device memory). Table~\ref{tab:CONV_mem_steps} reports similar qualitative results for \CONV models, which in our setup are composed by four large layers (all except the first one, which instead of $f$ input channels receives only 3).

\begin{table}[t]
\centering
\caption{Memory usage and inference time before and after each step for \FC models}
\label{tab:FC_mem_steps}
\resizebox{0.42\textwidth}{!}{%
\begin{tabular}{|c|c|c|c|c|}
\hline
\rowcolor[HTML]{EFEFEF} 
Step               & \#MACs   & Device (MiB) & Host (MiB) & Inf. time (ms)  \\ \hline
                    & 0.76e7 & 7.43 & 0 & 0.17  \\
\multirow{-2}{*}{1} & 0.79e7 & 5.27 & 2.63 & 7.42 \\ \hline
                    & 1.19e7 & 7.66 & 3.82 & 10.62 \\
\multirow{-2}{*}{2} & 1.24e7 & 4.04 & 8.04 & 21.83 \\ \hline
\end{tabular}%
}

\end{table}

\begin{table}[t]
\centering
\caption{Memory usage and inference time before and after each step for \CONV models}
\label{tab:CONV_mem_steps}
\resizebox{0.42\textwidth}{!}{%
\begin{tabular}{|c|c|c|c|c|}
\hline
\rowcolor[HTML]{EFEFEF} 
Step               & \#MACs  & Device (MiB) & Host (MiB) & Inf. time (ms) \\ \hline
                    & 2.88e10 & 6.86 & 0 & 41.34 \\
\multirow{-2}{*}{1} & 3.01e10 & 5.99 & 1.99 & 61.60 \\ \hline
                    & 3.87e10 & 6.78 & 2.25 & 69.71 \\
\multirow{-2}{*}{2} & 4.02e10 & 5.21 & 5.19 & 96.89 \\ \hline
                    & 5.89e10 & 6.98 & 6.95 & 126.41 \\
\multirow{-2}{*}{3} & 6.08e10 & 3.93 & 11.69 & 232.82 \\ \hline
\end{tabular}%
}

\vspace{-15pt}
\end{table}

The host memory overhead has a higher relative cost in \FC layers than in \CONV layers due to the arithmetic intensity difference between them. This is clearly seen when comparing the Edge TPU inference times with host inference times (see Figure~\ref{fig:EdgeTPUvsCPU_time}). Using the CPU, there are no communication overheads and the inference time is essentially increased by the amount of computation. For the \FC layers, we observe that the time difference between steps ($\sim$ 10 ms) is significantly higher than the CPU time of the slower models ($\sim$ 3 ms). In contrast, the time difference between steps for \CONV layers is negligible compared to the inference times on our CPU (especially if we limit the execution to a few cores). The Edge TPU stands out in these layers as the computation has a higher relative weight versus memory accesses, so the advantages of the systolic array versus more general processing are more noticeable. The difference is huge even though we are facing a low-end device against a high-end CPU.

\section{Model segmentation on multiple TPUs}\label{sec:multitpu}

To reduce host memory usage (and the associated penalty in performance), the compiler offers the option of segmenting the models to distribute the fragments among several Edge TPUs. 
The idea is to expand the effective device memory space by aggregating multiple TPUs so that less host memory is needed and host-to-device communications are reduced. To run inferences, the outputs of each segment are used as inputs to the TPU that contains the next one. Although these communications are also done through the host, they can be cost-effective because the number of intermediate outputs that are transmitted is smaller than the amount of the weights we avoid sending (for example, in an \FC layer that has $m$ inputs and $n$ outputs we have $m \cdot n$ weights). In this way, we form a pipeline of devices in which the inference of different inputs can be executed in parallel.

\begin{figure}[ht!]
    \centering
    \centering
    \includegraphics[width=\columnwidth]{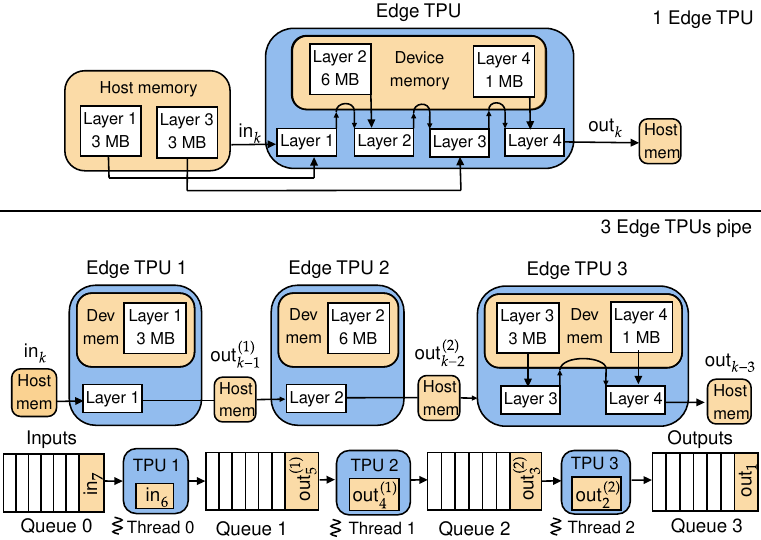}
    \caption{Top: Single-TPU execution for a segmented model with layers stored in the host and device memory. Bottom: Implemented pipelined execution scheme.}
    \label{fig:segmentation_TPUs_example}
\end{figure}

For model segmentation, we evenly distribute the layers between the desired segments. For example, our 5-layer models will be distributed among 3 segments with 1 layer for one of them and 2 layers for the remaining two. Obviously, the layers for each segment must be consecutive in the original model. 
%
To implement the pipelined execution, we deploy a host thread per Edge TPU to handle it, and a queue (implementing thread-safe Python mechanisms) on the host to communicate intermediate results among devices. Figure~\ref{fig:segmentation_TPUs_example} shows an example of a single TPU model run versus a 3-TPU segmentation illustrating a possible scenario in our pipelined implementation.

\begin{figure}[ht!]
    \centering
        \includegraphics[width=0.7\columnwidth]{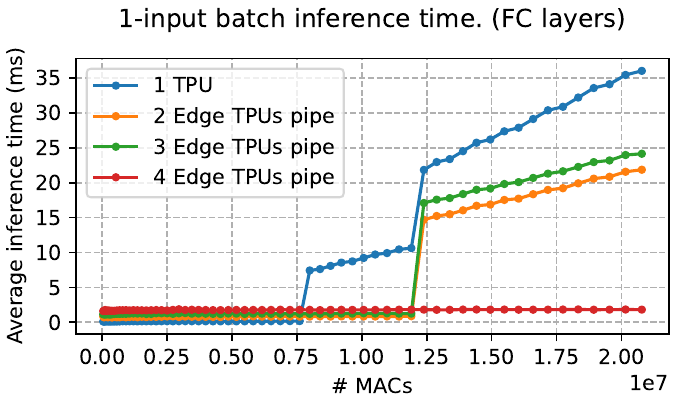}
        \includegraphics[width=0.7\columnwidth]{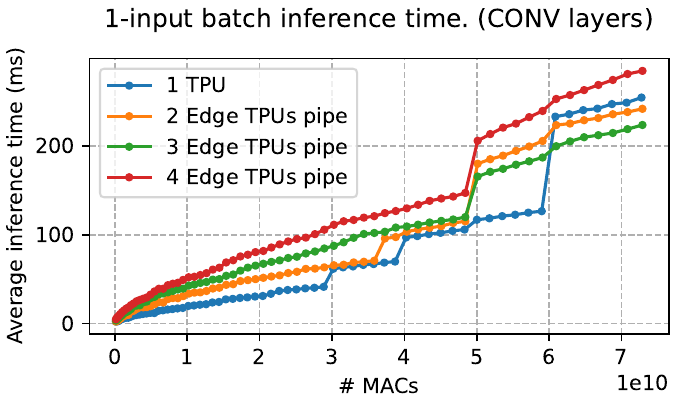}
    \caption{Inference time for \FC models (top) and \CONV models (bottom) on multiple TPUs using a pipelined implementation.}
    \label{fig:TPUs_pipeline_batch1}
\end{figure}

\subsection{Performance on single inputs with segmented models}\label{sec:performance:bs1}

We start by evaluating the inference time with different number of segments for a batch with a single input (see Figure~\ref{fig:TPUs_pipeline_batch1}). In this way, there is no parallelism in the execution and we will observe the positive impact of the reduction in communication caused by a more reduced usage of host memory combined with the negative impact of communicating TPUs among them.

The results with \FC layers are (qualitatively) much better than with \CONV layers since the communications we manage to reduce havea much higher relative weight in their inference time. In fact, we see that \FC models using host memory improve significantly with segmentation compared to their single-TPU counterpart. During the first step we can see the costs of communicating the TPUs (the times are slightly higher when more TPUsare used), which are practically negligible compared with the difference between steps.

In \CONV layers we did not expect as much improvement as in \FC layers because the overhead of loading host weights has a much lower relative cost. However, the reality is even worse because the ratio between the size of the intermediate tensors and the size of a layer is worse. This is because the convolution filters are moved over the input matrix and each weight is used over many input values. In Figure~\ref{fig:TPUs_pipeline_batch1} we observe that segmented runs are clearly slower than the single-TPU execution (except for the largest models which improve very slightly). Storing less layers in the host does not trade off for the communication of intermediate tensors between devices. Although the reuse of weights makes segmentation inefficient, we must remember that it is also the reason why the use of host memory is not so important in these layers.

For both layers, using more segments delays the need to store layers on the host lengthening the steps in our plots. However, the higher the number of segments, the higher the communication cost. The ideal is to use the minimum number of segments so that the model only uses device memories (it is in the first step). For example, in the case of \FC models with $\sim 10^7$ MAC operations, segmenting into two TPUs is sufficient to store the entire model in local memory and is better than segmenting into three or four TPUs. However, when the model is around $1.5 \cdot 10^7$ operations, using two or even three TPUs is not enough and it is worth using four TPUs for it. 

It is quite striking that \FC layer segmentations with two and three TPUs behave in the same way. A priori, with three TPUs the steps should be longer than with two. Moreover, with an ideal utilisation of the device memory, the three steps we observed in our \FC models should be reduced to one; however, four TPUs are needed. Similarly, the five steps that occurred in the convolution models could be reduced to one with four segments, but with that number of segments there are still two steps. To analyse these issues, we look at the memory usage data with multiple segments (see Tables~\ref{tab:mem_use_multisegment_FC} and ~\ref{tab:mem_use_4segment_CONV}).

We observe that both cases are a consequence of the uniform distribution strategy of the number of layers. In 3-TPUs segmentation for \FC models, the first chip only stores the first layer that barely has $64n$ weights (negligible compared to $n^2$ of other layers). For this reason, its device memory is practically not used and the segmentation behaves the same as with one less TPU. In 4-TPUs segmentation for \CONV models happens exactly the same. The first TPU only has a small layer, but the fourth TPU stores two large layers and ends up needing to use host memory. Ideally, the two layers stored on the same chip would be the first two (one large and one small).

\setlength{\tabcolsep}{2pt}

\begin{table}[t]
\caption{Memory usage of \FC models with 2 and 3 segments}
\label{tab:mem_use_multisegment_FC}

\resizebox{0.5\textwidth}{!}{%
\begin{tabular}{cc|cccc|cccccc|}
\cline{3-12}
                           &        & \multicolumn{4}{c|}{2 Edge TPUs (MiB)} & \multicolumn{6}{c|}{3 Edge TPUs (MiB)} \\ \hline
\rowcolor[HTML]{EFEFEF} 
\multicolumn{1}{|c|}{\cellcolor[HTML]{EFEFEF}n} &
  \cellcolor[HTML]{EFEFEF}\#MACs &
  \cellcolor[HTML]{EFEFEF}Dev 1 &
  \cellcolor[HTML]{EFEFEF}Dev 2 &
  Host 1 &
  Host 2 &
  Dev 1 &
  Dev 2 &
  Dev 3 &
  Host 1 &
  Host 2 &
  Host 3 \\ \hline
\multicolumn{1}{|c|}{1140} & 0.40e7 & 1.32        & 2.57        & 0        & 0          & 0.07    & 2.5     & 1.32    & 0   & 0      & 0   \\
\multicolumn{1}{|c|}{1380} & 0.58e7 & 1.94        & 3.79        & 0        & 0          & 0.09    & 3.71    & 1.94    & 0   & 0      & 0   \\
\multicolumn{1}{|c|}{1620} & 0.80e7 & 2.67        & 5.24        & 0        & 0          & 0.10    & 5.14    & 2.67    & 0   & 0      & 0   \\
\multicolumn{1}{|c|}{1860} & 1.05e7 & 3.52        & 6.93        & 0        & 0          & 0.12    & 6.81    & 3.52    & 0   & 0      & 0   \\
\multicolumn{1}{|c|}{2100} & 1.33e7 & 4.36        & 4.36        & 0        & 4.23       & 0.13    & 4.23    & 4.36    & 0   & 4.23   & 0   \\
\multicolumn{1}{|c|}{2340} & 1.65e7 & 5.43        & 5.43        & 0        & 5.28       & 0.14    & 5.28    & 5.43    & 0   & 5.28   & 0   \\
\multicolumn{1}{|c|}{2580} & 2.01e7 & 6.62        & 6.95        & 0        & 6.46       & 0.16    & 6.48    & 6.61    & 0   & 6.46   & 0   \\ \hline
\end{tabular}%
}

\end{table}

\setlength{\tabcolsep}{6pt}

\setlength{\tabcolsep}{2pt}
\begin{table}[t]
\centering
\caption{Memory usage of \CONV models with 4 segments}
\label{tab:mem_use_4segment_CONV}
\resizebox{0.43\textwidth}{!}{%
\begin{tabular}{cc|cccccccc|}
\cline{3-10}
                          &         & \multicolumn{8}{c|}{4 Edge TPUs (MiB)} \\ \hline
\rowcolor[HTML]{EFEFEF} 
\multicolumn{1}{|c|}{\cellcolor[HTML]{EFEFEF}f} &
  \cellcolor[HTML]{EFEFEF}\#MACs &
  Dev 1 &
  Dev 2 &
  Dev 3 &
  \cellcolor[HTML]{EFEFEF}Chip 4 &
  Host 1 &
  Host 2 &
  Host 3 &
  \cellcolor[HTML]{EFEFEF}Host 4 \\ \hline
\multicolumn{1}{|c|}{292} & 1.26e10 & 0.013  & 0.80  & 0.80  & 1.61 & 0 & 0 & 0 & 0    \\
\multicolumn{1}{|c|}{352} & 1.83e10 & 0.016  & 1.16  & 1.16  & 2.33 & 0 & 0 & 0 & 0    \\
\multicolumn{1}{|c|}{412} & 2.51e10 & 0.018  & 1.59  & 1.59  & 3.18 & 0 & 0 & 0 & 0    \\
\multicolumn{1}{|c|}{472} & 3.30e10 & 0.021  & 2.08  & 2.08  & 4.16 & 0 & 0 & 0 & 0    \\
\multicolumn{1}{|c|}{532} & 4.19e10 & 0.024  & 2.63  & 2.63  & 5.27 & 0 & 0 & 0 & 0    \\
\multicolumn{1}{|c|}{592} & 5.19e10 & 0.026  & 3.26  & 3.26  & 3.26 & 0 & 0 & 0 & 3.26 \\
\multicolumn{1}{|c|}{652} & 6.29e10 & 0.029  & 3.95  & 3.95  & 3.95 & 0 & 0 & 0 & 3.95 \\ \hline
\end{tabular}%
}
\vspace{-15pt}
\end{table}

\setlength{\tabcolsep}{6pt}

Based on these results, it seems logical to develop a partitioning that attempts to equalise memory usage between the segments. However, this solution would not consider that, for similar host memory usage, the one that distributes the workload more evenly is preferable. Ideally, the phases of our pipeline should have similar latency since performance will be limited by the slowest segment. In this sense, Google offers a profiling tool that tests the latency of the fragments for different distributions and tries to minimise the difference between the fastest and the slowest one. We will shortly analyse a profiling-based split, but let us first test the default segmentation scheme with a larger input batch.

\subsection{Performance on batched inputs with segmented models}\label{sec:performance:bsn}

To exploit the parallel potential of the pipeline, we repeated the experiment with a 50-input batch. In this case, we divide the execution time of the whole batch by its size to obtain the time per inference. Figure~\ref{fig:TPUs_pipeline_batch50} reports the speedup with respect to a single input and with respect to one TPU.

In both layer types, the speedup with respect to a single input is far from ideal (with $n$ TPUs we could expect close to $\times n$). The problem is that the workload distribution is unbalanced and there are stages much slower than others acting as bottlenecks. When the model fits in device memory the speedup is moderately below ideal because the workload distribution is uneven. When a TPU also needs host memory, it becomes a very slow stage in the pipeline that sequences the executions. Therefore, the speedup with respect to a single input drops sharply near $\times 1$ when host memory is needed.

These load imbalance issues combined with the communication overhead make speedups compared to a single TPU very poor for \CONV models and for \FC models that fit on-chip or still need host memory. Recall that in \CONV layers the communication costs are very relevant on their own and it is natural that, despite parallelisation, segmentation is inefficient (in many models it is still slower than 1 TPU). However, the results are very positive for \FC models that avoid the use of host memory completely (because its relative cost is very high). In these cases, we obtain speedups of several tens with just 2, 3 or 4 TPUs compared to using just one. We get up to $\times 36$ in the largest models (with just 4 TPUs) where the size of the layers is already too big to be stored on the host.

\subsection{Optimizing model segmentation with profiling}\label{sec:performance:profiling}

We have observed two different issues to improve segmentation through a better model distribution. First, it is convenient to make a better use of device memory with a more balanced distribution of layers in terms of their size. In this way, we will achieve a further reduction in host memory usage for the same number of devices. Second, it is desirable for the workload to be more evenly distributed so that the pipeline stages exhibit similar latencies and hence the load imbalance is reduced.

In our synthetic models, both purposes match, since all layers are of the same type (they are all either \FC or \CONV). As they all exhibit the same arithmetic intensity, a higher workload entails a higher memory usage. However, it is very common for models to combine layers of different types where some take up more than others in relation to the amount of work they perform. In this sense, simultaneously tackling both aspects by using parameters such as the amount of memory and the number of MAC operations would require a multivariable optimisation problem that is expensive to solve. Instead, we propose to profile the execution of different partitions in order to choose which one to use.

\begin{figure*}[t]
    \centering
    \hspace{-23pt}
    \begin{minipage}[t]{0.66\textwidth}
        \begin{minipage}[c]{0.5\textwidth}
        \centering
        \subfloat{
            \includegraphics[width=\textwidth]{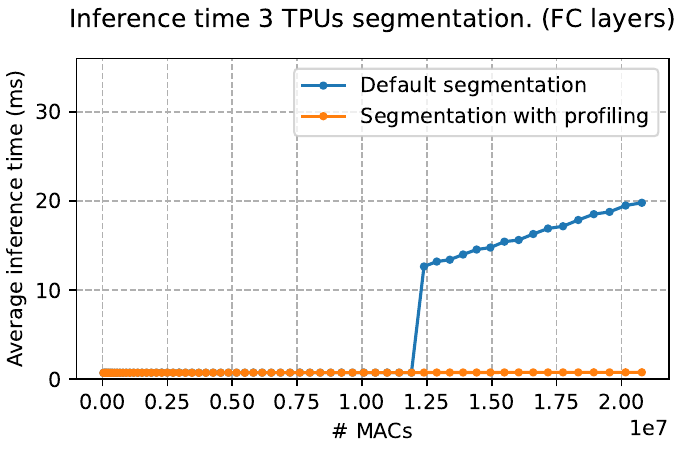}
        }\\
        \subfloat{
            \includegraphics[width=\textwidth]{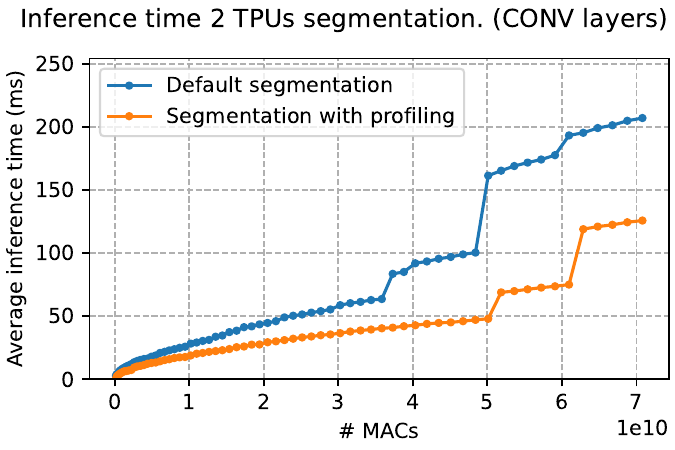}
        }
        \end{minipage}
        \begin{minipage}[c]{0.5\textwidth}
        \centering
        \subfloat{
          \includegraphics[width=\textwidth]{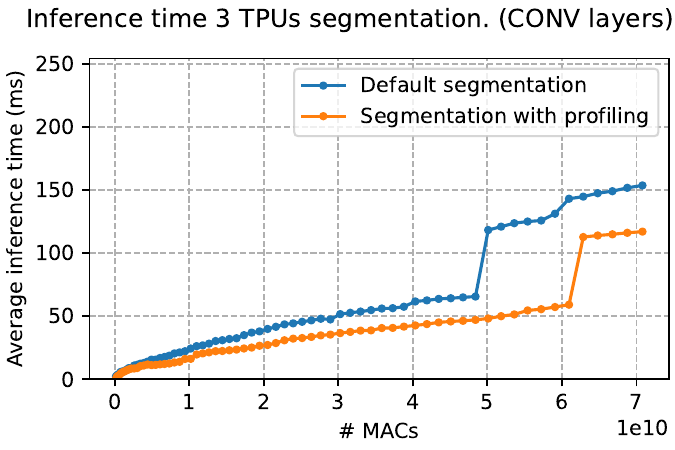}
        }\\
        \subfloat{
          \includegraphics[width=\textwidth]{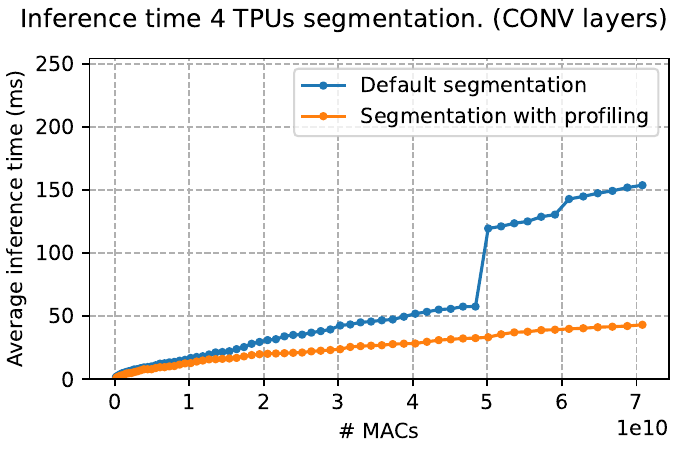}
        }
        \end{minipage}
    \caption{Inference time for segmented models with profiling for \FC and \CONV layers using multiple TPUs for a 50-input batch.}
    \label{fig:time_inference_profiling}
    \end{minipage}
    \hspace{5pt}
    \begin{minipage}[t]{0.33\textwidth}
    \centering
        \subfloat{
            \includegraphics[width=\textwidth]{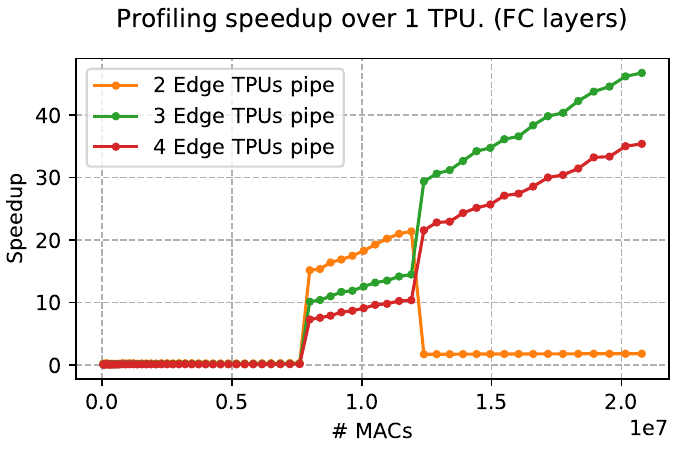}
        }\\
        \subfloat{
            \includegraphics[width=\textwidth]{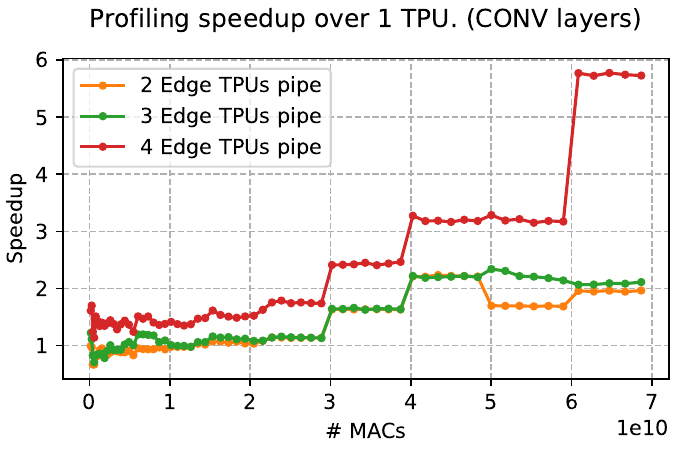}
        }
         \caption{Speedups of segmented models for \FC (top) and \CONV (bottom).}
        \label{fig:speedup_vs_1TPU_profiling}
    \end{minipage}
\end{figure*}

Google offers a compilation tool that profiles the segmentation of different partitions for the number of segments specified by the user\footnote{\url{https://coral.ai/docs/edgetpu/compiler/\#profiling-partitioner}}. The profiling tool receives a parameter with the desired maximum difference between the time of the fastest segment and that for the slowest segment. Different partitions are tested until one meets the constraint and is therefore chosen. In case no schedule meets the requested constraint, the last tested configuration is chosen. We assume that the profiling tool from Google approaches profiling with an objective threshold to avoid compilation becoming too slow by trying all options if the user's constraints are undemanding.

However, if the model has few layers, the number of possible distributions is affordable for exhaustive exploration: there are $(l-1)!/\left((s-1)!\cdot(l-s)!\right)$ possibilities\footnote{It is about splitting the $l$ layers into $s$ segments. This is equivalent to choosing $s-1$ separators (to form the $s$ segments) among the $l-1$ positions between layers. That is, the possibilities are $\binom{l-1}{s-1} = \frac{(l-1)!}{(s-1)!\cdot(l-s)!}$.}, where $l$ is the number of layers and $s$ is the number of segments. In our $5$ layer models there are only $14$ different possibilities. Therefore, we implemented exhaustive profiling exploring all options and taking the best one in terms of the inference time when running as a pipeline over a large batch of inputs.

In Tables~\ref{tab:mem_use_multisegment_FC} and \ref{tab:mem_use_4segment_CONV} we discovered scenarios where the compiler's default pipeline underuses the memory of one TPU while some layers had to be stored on the host because the capacity of the device memory was exceeded. In contrast, Tables~\ref{tab:FC_3TPUs_pipeline_profiling} and \ref{tab:CONV_4TPUs_pipeline_profiling} reveal that profiling-based segmentation is more equitable in memory usage. In the case of 3 TPUs for \FC layers, the first TPU takes a large layer that was stored by the second TPU in the default segmentation. For 4 TPUs (with \CONV layers), the same happens for a layer that was saved in the fourth TPU by default. In fact, this partitioning manages to store all models completely in device memories as we already predicted could happen with that number of devices.

Figure~\ref{fig:time_inference_profiling} reveals that the inference time steps are longer with profiling as a consequence of a more balanced distribution of the model size. The cases of \FC layers with 2 and 4 TPUs are not shown because the profiling chooses the same segmentation as the default. In the cases where there is a better segmentation, one less layer is kept in the host for many models (they are in one step less than in the default segmentation) and the inference time is reduced considerably. Furthermore, in \CONV layers a reduction in times is also observed due to a more equal distribution of the workload that balances the latency of the segments. Inference times during the first step (host memory has not yet been used) are lower with profiling than without profiling. As the workload is more balanced, the pipeline stages have similar latency and parallelization is more efficient. In the case of \FC layers it is negligible because their workloads are still too reduced (especially compared to the cost of storing a 
layer on the host).

Finally, in Figure~\ref{fig:speedup_vs_1TPU_profiling} we analyze the speedup of a parallel execution with profiling compared with the initial case of a single TPU. We have already observed that the default segmentation was very profitable in \FC layers when it avoided using host memory altogether. With profiling, this situation is extended to all partitions for three TPUs, which is the best option for models that would save two layers on the host without segmentation. Although four TPUs also avoid using host memory, their results are worse due to the unnecessary communication overhead of an extra TPU. The same happens in the models that would only store one layer in the host, whose maximum acceleration is obtained with two TPUs. Definitely, the optimum is to use the minimum number of TPUs that avoids using host memory.

In \CONV layers, profiling improves the default segmentation to the point where it is slightly cost-effective to segment larger models (with 4 TPUs, we avoid storing 3 layers in host memory and get about $\times 6$ over one TPU). This happens despite the fact that using host memory does not have as much impact as in the \FC layers and the communication costs between TPUs are relatively high in relation to the workloads. However, in models that with a TPU would only save one or two layers, it is not worth investing the hardware in segmenting. Alternative strategies, such as replicating the model (i.e., model parallelism) and partitioning the input batch (i.e., data paralellism), might be more efficient.

\section{Conclusions}\label{sec:conclusions}


In this paper we have analyzed the performance of the Edge TPU for different types of neural networks, comprising both \FC and \CONV layers. After identifying the main performance penalties in single-TPU executions (mainly due to the use external memory for weight allocation), we have proposed the use of profiled model segmentation and pipelining techniques not only to reduce the execution time by providing parallelism across TPUs, but also to remove the burden introduced by remote memory accesses. The experimental results reveal remarkable peroformance improvements ranging from $6\times$ for \CONV models, to $46\times$ for \FC models.

Even though the tested models are synthetic, our conclusions are general enough to help the programmer to understand the behavior of more complex models, possibly with heterogeneous layers both in type and number of nodes/filters. Future work will include hybrid CPU-TPU inference executions following similar pipelined implementations, a deeper study on the energy efficiency of single- and multi-TPU implementations and evaluation of the trade-offs between performance, precision and energy efficiency of this type of accelerators.

\section*{Acknowledgement}
This work has been partially supported by Grant PID2021-126576NB-I00 funded by MCIN/AEI/ 10.13039/501100011033 and by “ERDF A way of making Europe”, and the CM under Grant S2018/TCS-4423.

\bibliographystyle{IEEEtran}
\bibliography{biblio}

\end{document}